%% file: mypaper.tex
\documentclass[conference]{IEEEtran}
\IEEEoverridecommandlockouts

\usepackage{amsmath,amsfonts}
\usepackage{algorithmic}
\usepackage{algorithm}
\usepackage{array}
\usepackage{textcomp}
\usepackage{stfloats}
\usepackage{url}
\usepackage{verbatim}
\usepackage{graphicx}
\usepackage{subfigure}
\usepackage{cite}
\usepackage{upgreek}
\usepackage{amssymb} 
\usepackage[justification=centering]{caption}
\usepackage{subcaption}
\newtheorem{theorem}{Theorem}
\newtheorem{lemma}{Lemma}

\usepackage{xcolor}
\def\BibTeX{{\rm B\kern-.05em{\sc i\kern-.025em b}\kern-.08em
    T\kern-.1667em\lower.7ex\hbox{E}\kern-.125emX}}
\begin{document}

\title{Multi-Cell Coordinated Beamforming for Integrate    Communication and Multi-TMT Localization
}

\author{
    \IEEEauthorblockN{Meidong Xia$^{1}$, Wei Xu$^{1}$,  Jindan Xu$^{2}$, Zhenyao He$^{1}$, Zhaohui Yang$^{3}$, and Derrick Wing Kwan Ng$^{4}$}
    \IEEEauthorblockA{$^1$National Mobile Communications Research Laboratory, Southeast University, Nanjing 210096, China}
    \IEEEauthorblockA{$^2$School of Electrical and Electronics Engineering, Nanyang Technological University, Singapore 639798, Singapore}
    \IEEEauthorblockA{$^3$College of Information Science and Electronic Engineering, Zhejiang University, Hangzhou 310027, China}
    \IEEEauthorblockA{$^4$School of Electrical Engineering and Telecommunications, University of New South Wales, Sydney, NSW 2052, Australia}
    \IEEEauthorblockA{Emails: \{meidong.xia, wxu, hezhenyao\}@seu.edu.cn, jindan.xu@ntu.edu.sg, yang\_zhaohui@zju.edu.cn, w.k.ng@unsw.edu.au}

    \thanks{This work was supported in part by the National Key Research and 
            Development Program under Grant 2024YFA1014204, and the Special Fund for 
            Key Basic Research in Jiangsu Province No. BK20243015. (Corresponding author: Wei Xu)}
    }
\maketitle

\input{./subText/abstract.tex}

\input{./subText/keywords.tex}

\input{./subText/introduction.tex}

\input{./subText/systemmodel.tex}

\input{./subText/digital.tex}

\input{./subText/simulation.tex}
\input{./subText/Conclusion.tex}

\input{./subText/Appendix.tex}

\bibliographystyle{IEEEtran}
\bibliography{ref.bib}

\end{document}

%% file: subText/abstract.tex
\begin{abstract}
	This paper  investigates   integrated localization and communication in a multi-cell system,
	and proposes a coordinated beamforming algorithm to enhance   target localization accuracy
	while preserving communication performance. 
	Within this integrated sensing and communication (ISAC) system, the Cramér-Rao lower bound (CRLB) is adopted to quantify  
	the accuracy of target localization, 
	with  its closed-form expression   derived   for the first time.
	It is shown that the nuisance parameters can be disregarded without impacting the CRLB of  
	time of arrival (TOA)-based target localization.
	Capitalizing on 
	the derived CRLB, 
	we formulate a nonconvex coordinated beamforming  problem  to minimize
	the CRLB  while satisfying   
	signal-to-interference-plus-noise ratio (SINR) constraints in communication. 
	To facilitate the development of   solution,  
	we  reformulate the original  problem into a more tractable form and solve it
	through semi-definite programming (SDP).
	Notably, we show that the proposed algorithm
	can always obtain  rank-one global optimal solutions under mild conditions.
	Finally, numerical results demonstrate the superiority of the proposed algorithm over benchmark algorithms
	and reveal the  performance trade-off 
	between localization accuracy and communication SINR.
\end{abstract}

%% file: subText/keywords.tex

%% file: subText/Introduction.tex
\section{Introduction}
In recent years, 
emerging applications have necessitated 
the next-generation wireless communication systems
to be more flexible  and intelligent
than   current offerings  \cite{xuTowardsUbiquitous2023,xuEdgeLearningB5G2023,
XuDisentangledRepresentationLearning2024}.
One promising paradigm for achieving these goals in
next-generation wireless systems is
integrated sensing and communication (ISAC) \cite{ZTE}.
Compared to deploying separate sensing and communication, an integration of
sensing and communication functionalities enables the system 
to share resources such as antennas  and power, which significantly reduces   deployment costs 
and improves   system efficiency \cite{liuIntegratedSensingCommunications2022,SturmWaveformDesignSignal2011}.

To unleash its potential, 
intensive studies have been conducted on ISAC systems with a single base station (BS).
A key area of research focuses  on   transmission 
design aimed at enhancing  both communication-related  
and sensing-related metrics. 
For instance, beampattern matching error and signal-to-interference-plus-noise ratio (SINR) are the key design criterions 
in  
\cite{liuJointTransmitBeamforming2020}, striving to  
achieve  desired sensing performance
while guaranteeing   communication quality of service (QoS).
The authors in \cite{heFullDuplexCommunicationISAC2023a} investigate 
an ISAC system based on full-duplex communication. They demonstrate that 
optimal beamforming design can simultaneously achieve the desired sensing SINR 
and improve the communication rate.
Furthermore, 
a perceptive mobile network (PMN) architecture with target 
monitoring terminals (TMTs) is proposed in \cite{xiePerceptiveMobileNetwork2022}, 
where the reception of sensing signals 
is delegated to the TMTs. 

Despite the fruitful research results in literature,
single-BS ISAC systems often face challenges 
in terms of capacity  
and accuracy.
To overcome these challenges, 
multi-BS ISAC systems have attracted increasing attention.
One promising
framework  for these systems is the multi-cell MIMO with interference coordination.
In particular, existing researches on multi-cell ISAC systems mainly focus on the  
scenarios of target detection \cite{chengOptimalCoordinatedTransmit2024} and parameter 
estimation \cite{chenFastFractionalProgramming2024}.
When it comes to the target localization, 
there is limited research.
The authors in
\cite{wangConstrainedUtilityMaximization2021} and \cite{huangCoordinatedPowerControl2022}
consider the target localization scenario in multi-cell ISAC systems,
but they all  investigate the power control problem and focus on   single-antenna BS configuration.
This 
restriction hampers
both sensing  accuracy and communication capacity. 

Compared to   power control in single-antenna systems,  beamforming design
in multi-antenna networks is more challenging.
On the one hand, the presence of more nuisance parameters (e.g., the angles)
makes the derivation of Cramér-Rao lower bound (CRLB) for location estimation more intricate.
It is noteworthy that  alternative   metrics, such as the beampattern matching error and the sensing SINR,
can also be adopted  as   design metrics 
for target localization,  
but they are not directly related to   target localization accuracy 
in the same way as the CRLB is.
On the other hand, the more complex expression of CRLB for location estimation 
also makes the beamforming design problem   harder to solve.
To the best of our knowledge, these challenges have not been fully investigated in the multi-cell integrated localization and communication
systems with  multi-antenna BS configuration.

In this paper, a multi-cell integrated localization and communication system 
is considered.
We derive the closed-form expression of CRLB for location estimation
in this   system and then formulate a coordinated beamforming optimization problem
to minimize the CRLB while satisfying   communication SINR constraints.
We first transform  
the fractional structure of the problem into two matrix inequalities and then solve it
by  semi-definite programming (SDP).
Simulation results verify the effectiveness of the proposed algorithm in enhancing the localization accuracy
compared to benchmark algorithms.

\subsubsection*{Notations}
In this paper, 
we use $\left\| \cdot \right\|$, $\operatorname{tr}\left(\cdot\right)$, $\left( \cdot \right)^T$, $\left( \cdot \right)^H$, 
and $\left( \cdot \right)^*$ to represent the Euclidean norm,
the trace operation, the transpose, 
the Hermitian transpose, and the complex conjugate of   matrices or vectors, respectively.
$\mathrm{diag}\left( \cdot \right)$ denotes the diagonal matrix operator.
$\mathbb{C}$ and $\jmath$ represent the set of complex numbers and the imaginary unit, respectively.
$\partial$ represents the partial derivative.
$\left\{ \mathbf{A}_{m,k} \right\}$ represents the set of all elements $\mathbf{A}_{m,k}$ 
for $m=1,2,\ldots,M$ and $k=1,2,\ldots,K$.
$\left[\mathbf{A}\right]_{m,n}$ denotes the element in the $m$-th row and $n$-th column of matrix $\mathbf{A}$.
$\mathrm{span} \left( \cdot \right)$ denotes the span of a set of vectors, while 
$\bigcup$ denotes the union of sets.
Furthermore, $\mathrm{Re}\left( \cdot \right)$ 
is used to denote the real parts  
of a complex scalar.

%% file: subText/systemmodel.tex
\section{System Model and Problem Formulation}
\begin{figure}[t]
	\centering
    \includegraphics[width=0.43\textwidth]{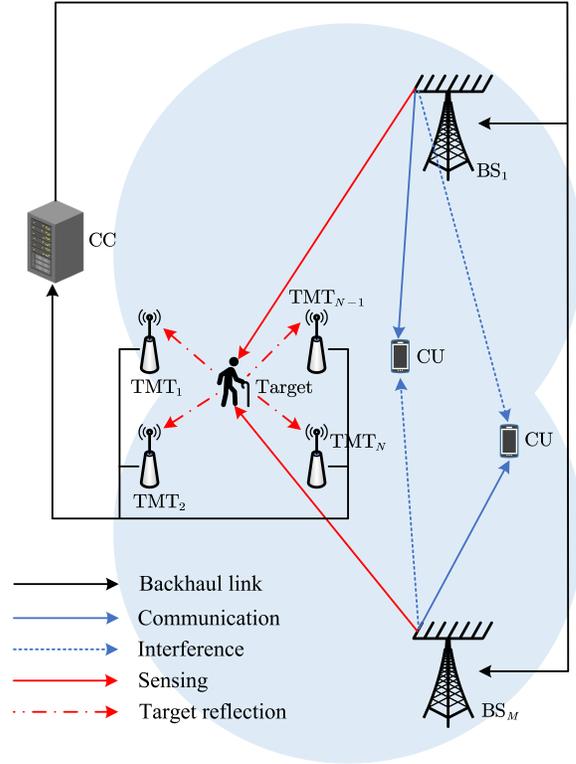}
	\caption{System model of the multi-cell ISAC system.}
	\label{figsetup}
\end{figure}

As shown in Fig. \ref{figsetup}, we consider a multi-cell ISAC system  including $M$ BSs, 
$N$ TMTs, 
a central controller (CC), 
$K$ single-antenna communication users (CUs) per BS, and a target to be sensed,
where each BS is equipped with $N_{\mathrm{t}}$ transmit antennas and each TMT possesses 
a   receiving antenna. 
The  BSs   are connected 
to the CC via backhaul links and  
transmit the ISAC signals to 
their associated CUs and
the target.
The TMTs  are  deployed in the target area and are
responsible for collecting and transferring the received signals to the CC for further processing.
The CC is responsible for coordinating the ISAC signal transmission 
and for processing the sensing signals to estimate the target location. 
\subsection{Transmit Model}
The signal sent by the $m$-th BS to its $k$-th CU is expressed as
\begin{equation}
    s_{m,k}(t)=\sum_{l=1}^{L} c_{m,k,l}g(t-lT_{\mathrm{s}}), \ \forall m, k.
\end{equation}
Here, $L$, 
$T_{\mathrm{s}}$, and $c_{m,k,l}$ denotes the number of sensing  snapshots, 
the duration of each symbol,  and the   transmitted symbol
of  the $m$-th BS for  serving the $k$-th user 
at the $l$-th snapshot, respectively.
Also, 
$g(t)$ denotes the real-valued 
baseband pulse signal which satisfying 
\begin{equation}
    \int_{-\infty}^{\infty} f\left|G(f)\right|^{2}\mathrm{d}f=0, \label{eq:real}
\end{equation} 
where $G(f)$ is the 
Fourier transform of $g(t)$. 
Assume that $L$ is sufficiently large, then the transmitted signals are unit-power and orthogonal to each other,
i.e.,
\begin{subequations}
    \begin{align}
        \frac{1}{L T_{\mathrm{s}}}\int_0^{L T_{\mathrm{s}}} \left| s_{m , i}(t) \right|^2  \mathrm{d} 
        t&=1, \  \forall m, i, \\
        \int_0^{L T_{\mathrm{s}}} s_{m , i}(t) s_{n, j}^*(t-\tau) \mathrm{d} 
        t&=0, \ \forall \tau, \left\{m,i\right\} \neq \left\{n,j\right\}. 
    \end{align}
\end{subequations}
Let $\mathbf{f}_{m,k} \in \mathbb{C}^{N_{\mathrm{t}} \times 1}$  denotes the $k$-th beamforming vector associated 
with  the $m$-th BS. In this work, we operate under the assumption that the  transmit power of each BS
is constrained by a predefined limit, denoted by $P$. 
Mathematically, this constraint is expressed as
\begin{equation}
    \sum_{k=1}^{K} \left\|\mathbf{f}_{m, k}\right\|^{2} \leq P, \ \forall m.
\end{equation}

\subsection{Communication Model}
Let us denote the channel from 
the $n$-th BS to the $k$-th user  served by the $m$-th BS  as $\mathbf{h}_{n, m, k} \in \mathbb{C}^{N_{\mathrm{t}} \times 1}$.
Then, the  signal that the $k$-th user  served by the $m$-th BS  receives at time instance $t$ is given by
\begin{equation}
    \begin{aligned}
        \hat{s}_{m,k}(t) & =  \mathbf{h}_{m, m, k}^H \mathbf{f}_{m, k} s_{m,k}(t) \\
        &  +  
        \sum_{j=1, j \neq k}^{K }\mathbf{h}_{m, m, k}^{H} \mathbf{f}_{m, j} s_{m, j}(t) \\
        & + \sum_{i=1, i \neq m}^{M} \sum_{j=1}^K \mathbf{h}_{i, m, k}^H \mathbf{f}_{i, j} s_{i, j}(t)  + z_{m,k}(t),
    \end{aligned} \label{5}
\end{equation} 
where $z_{m,k}(t)$ is the circularly symmetric complex Gaussian (CSCG) noise, 
characterized by a  zero mean and  a variance denoted by $\sigma_{\mathrm{n}}^2$.
Without loss of generality,  the SINR
is chosen as the performance metric for  communication. From $\eqref{5}$,  the SINR
of the $k$-th user served by the $m$-th BS  is given by $\eqref{eq-sinr}$ at the top of next page.

\begin{figure*}
    \begin{equation}
        \label{eq-sinr}
        \mathrm{SINR}_{m, k}=\frac{\left|\mathbf{h}_{m, m, k}^H 
        \mathbf{f}_{m, k}\right|^2}{\sum_{j=1, j \neq k}^{K }\left|\mathbf{h}_{m, m, k}^{H}
        \mathbf{f}_{m, j}\right|^2+\sum_{i=1, i \neq m}^{M} \sum_{j=1}^K\left|\mathbf{h}_{i, m, k}^H \mathbf{f}_{i, j}\right|^2+
        \sigma_{\mathrm{n}}^2}
    \end{equation}
    \hrulefill
\end{figure*}

\subsection{Sensing Model}
At a given time instance $t$, the ISAC signal received by the $n$-th TMT
can be mathematically expressed as follows
\begin{equation}
    \begin{aligned}
        y_n(t)    =  
        \underbrace{\sum_{m=1}^M   \sum_{k=1}^K \varepsilon_{m, n}  
        \mathbf{a}^T   \left(\theta_m\right)   \mathbf{f}_{ m, k}  s_{m, k}  \left(t  -  \tau_{m, n}\right)}_{\mu_n(t)}    +    n_n(t).
    \end{aligned}
\end{equation}
Herein, $\tau_{m, n}$ represents the propagation delay of the ISAC 
signal from the $m$-th BS to the $n$-th TMT. 
Meanwhile, $\varepsilon_{m, n}$ refers to the channel coefficient, incorporating 
both the large-scale channel fading coefficient and the radar cross section (RCS). 
$\theta_m$ represents the angle of departure (AoD) from the $m$-th BS to the target.
$\mathbf{a}\left(  \cdot \right)$ denotes the array response  vector.
$\mu_n(t)$ and $n_n(t)$ represent the useful signal and CSCG noise, respectively. 
Notably, we neglect  the direct path between the BSs and the TMTs
in the considered  signal model for sensing, because the 
CC is assumed to have prior knowledge of the ISAC signals, 
enabling the estimation and removal of the  direct path before estimating the target's location.

The aggregation of signals at the CC can be mathematically represented as 
\begin{equation}
    \mathbf{y}(t)=\left[\begin{array}{c}
        {\mu}_1(t), {\mu}_2(t)
        \cdots, 
        {\mu}_{N}(t)
    \end{array}\right]^T+\mathbf{n}_{\mathrm{s}}(t),
\end{equation}
where $\mathbf{n}_{\mathrm{s}}(t)$ denotes the CSCG   
noise, which is   
characterized by a zero mean and an autocorrelation function $\sigma_{\mathrm{s}}^2 \mathbf{I}_N
\delta(\tau)$.  
$\sigma_{\mathrm{s}}^2$ and $\delta(\tau)$ denote the
power spectral density (PSD) of  noise and the Dirac delta function, respectively.
Thanks to the orthogonality 
among the transmitted  signals,
the CC   estimates the time delay  from variations
in the envelope of the transmitted signals.
Employing the well-known time of arrival (ToA) localization method, 
the location of  target is determined by   a set 
of equations if the time delays are estimated \cite{godrichTargetLocalizationAccuracy2010}, 
which are expressed as
\begin{equation}
    \label{tau}
    \begin{aligned}
    \tau_{m,n}&=\frac{1}{\mathrm{c}}\left(\sqrt{\left(x_{m}-x\right)^2+\left(y_{m}-y\right)^2}\right.  \\
    & \left.+\sqrt{\left(x^{\prime}_{n}-x\right)^2+\left(y^{\prime}_{n}-y\right)^2}\right), \ \forall m, n, 
    \end{aligned}
\end{equation}
where $\mathrm{c}$ represents the speed of light,  $x_{m}$ and $y_{m}$ denote the x-coordinate and the y-coordinate of the $m$-th BS, 
$x^{\prime}_{n}$ and $y^{\prime}_{n}$ signify the corresponding  coordinate   of the $n$-th TMT,  and
$x$ and $y$ correspond to the coordinate of the target, respectively. 

To assess  the accuracy of target localization, the CRLB is adopted as the  performance metric, 
as it provides the lower bound on the variance that unbiased estimators can achieve \cite{godrichTargetLocalizationAccuracy2010}.
In the following, we  derive the CRLB  for the estimation of  target location.

\begin{lemma}
	The sum of  CRLB for the ToA-based estimation of target location $\left(x,y\right)$ is given by
    \begin{equation}
        \label{CRLBxy}
        {C}_{x,y} = \left[\mathbf{C}\right]_{1,1} + \left[\mathbf{C}\right]_{2,2} =
        \mathrm{tr} \left( \left(\boldsymbol{\Lambda} \mathbf{Z} \boldsymbol{\Lambda}^T\right)^{-1} \right),
    \end{equation}
    where 
    \begin{equation}
        \setlength{\arraycolsep}{1pt}
        \begin{aligned}
        &\boldsymbol{\Lambda} =
        \left[\begin{array}{cccc}
        \frac{\partial}{\partial x} \boldsymbol{\uptau}^T \\
        \frac{\partial}{\partial y} \boldsymbol{\uptau}^T \\
        \end{array}\right]
        \end{aligned}
    \end{equation}
    denotes the Jacobian sub-matrix which is
    composed of the partial derivatives $\frac{\partial}{\partial x} \boldsymbol{\uptau}^T$ 
    and $\frac{\partial}{\partial y} \boldsymbol{\uptau}^T$, and
    \begin{equation}
        \label{Z}
        \begin{aligned}
            \mathbf{Z} = \mathrm{diag}& \left(J(\tau_{1,1},\tau_{1,1}),\ldots,J(\tau_{M,N},\tau_{M,N})\right)
        \end{aligned}
    \end{equation}
    signifies the Fisher information sub-matrix  related to the time delays parameters.
    $J(\tau_{m,n},\tau_{m,n})$ denotes the element   with respect to time delay $\tau_{m,n}$, 
    expressible as
    \begin{equation}
        \begin{aligned}
            J(\tau_{m,n},\tau_{m,n}) & = \frac{8\pi^2 L T_{\mathrm{s}}  \beta^2  \left| \varepsilon_{m,n} \right|^2}
            {\sigma_{\mathrm{s}}^2}  \\ 
            & \times  \mathbf{a}^T\left(\theta_m\right) 
            \left( 
            \sum_{k=1}^{K}\mathbf{f}_{m,k} \mathbf{f}_{m,k}^{H} \right)  \mathbf{a}^{*}\left(\theta_m\right),
            \label{J_ele}
        \end{aligned} 
    \end{equation}
    where $\beta$ is the effective bandwidth of the pulse signal $g(t)$.
\end{lemma}
\begin{IEEEproof}
    Let $\boldsymbol{\upphi} \triangleq \left[x, y, \boldsymbol{\uptheta}^T, {\boldsymbol{\upvarepsilon}_{\mathrm{R}}}^T, 
    {\boldsymbol{\upvarepsilon}_{\mathrm{I}}}^T\right]^T$ represent the parameter vector.
    Here, $\boldsymbol{\uptheta} \triangleq \left[\theta_{1}, \theta_{2}, \ldots, \theta_{M}\right]^T$, while 
    ${\boldsymbol{\upvarepsilon}_{\mathrm{R}}}$ and ${\boldsymbol{\upvarepsilon}_{\mathrm{I}}}$ respectively
    represent the real and imaginary parts  of  $\boldsymbol{\upvarepsilon} \triangleq \left[\varepsilon_{1,1}, \ldots,  \varepsilon_{M, N}\right]^T$.
    The parameters $\boldsymbol{\uptheta}$, ${\boldsymbol{\upvarepsilon}_{\mathrm{R}}}$, and ${\boldsymbol{\upvarepsilon}_{\mathrm{I}}}$ are considered as nuisance parameters,
    which are not directly related to the ToA-based target localization.
    Then, the CRLB with respect to $\boldsymbol{\upphi}$ is defined as  
    \cite{godrichTargetLocalizationAccuracy2010}
    \begin{equation}
        \label{CRLB}
        \mathbf{C}=\mathbf{J}^{-1}(\boldsymbol{\upphi}),
    \end{equation}
    where $\mathbf{J}(\boldsymbol{\upphi})$ denotes  the Fisher information  matrix 
    (FIM) pertaining to the parameter vector
    $\boldsymbol{\upphi}$.
    In this context of ToA-based localization, it is more straightforward to calculate the FIM
    when done for an alternative parameter vector $\boldsymbol{\uppsi}$. 
    Specifically, we define the parameter vector $\boldsymbol{\uppsi} \triangleq
     \left[\boldsymbol{\uptau}^T, \boldsymbol{\uptheta}^T, {\boldsymbol{\upvarepsilon}_{\mathrm{R}}}^T, 
    {\boldsymbol{\upvarepsilon}_{\mathrm{I}}}^T\right]^T$, with $\boldsymbol{\uptau} \triangleq \left[\tau_{1,1}, \ldots,   \tau_{M, N}\right]^T$.
    By the chain rule \cite{godrichTargetLocalizationAccuracy2010}, we have
    \begin{equation}
        \label{Jphi}
        \mathbf{J}(\boldsymbol{\upphi})=\mathbf{G} \mathbf{J}(\boldsymbol{\uppsi}) \mathbf{G}^T,
    \end{equation}
    where $\mathbf{J}(\boldsymbol{\uppsi})$ denotes the FIM relative to $\boldsymbol{\uppsi}$, and 
    $\mathbf{G}=\frac{\partial \boldsymbol{\uppsi}}{\partial \boldsymbol{\upphi}}$
    denotes the Jacobian matrix. 

    By selectively focusing on the
    components   that   affect the CRLB of target localization,
    the Jacobian matrix $\mathbf{G}$ can be   represented in the following structured form 
    \begin{equation}
        \label{G}
        \mathbf{G}  =\frac{\partial \boldsymbol{\uppsi}}{\partial \boldsymbol{\upphi}} = \left[  \begin{array}{cc}
        \mathbf{\Lambda} & \mathbf{0}_{2 \times (2MN+M)} \\
        \mathbf{0}_{(2MN+M) \times MN} & \mathbf{\Theta}
        \end{array} \right], 
    \end{equation}
    where $\mathbf{\Lambda} \in \mathbb{C}^{2 \times MN}$  is of paramount interest and
    can be calculated based on  $\eqref{tau}$.
    Nevertheless,  $\mathbf{\Theta}$ represents a Jacobian sub-matrix of no interest 
    for our current analysis, as it is related to the nuisance parameters and does not affect the CRLB of target localization.

    \begin{lemma} \label{lemma2}
        The analytical derivation of the FIM  $\mathbf{J}(\boldsymbol{\uppsi})$  
        is expressed in the structured form
        \begin{equation}
            \label{Jpsi}
            \mathbf{J}(\boldsymbol{\uppsi})  =  
            \begin{bmatrix}
                \mathbf{Z}  &  \mathbf{0}_{MN \times (2MN+M)} \\
                \mathbf{0}_{  (2MN+M) \times MN}  &   {\boldsymbol{\Omega}} \\
            \end{bmatrix},
        \end{equation}
        where $\mathbf{Z} \in \mathbb{C}^{MN \times MN}$ is described in $\eqref{Z}$ and $\eqref{J_ele}$, while
        $\mathbf{\boldsymbol{\Omega}}$ is the Fisher information sub-matrix
        with respect to nuisance parameters, which exerts no influence on 
        the CRLB of ToA localization.
    \end{lemma}

    \begin{IEEEproof}
        The proof is provided in Appendix A.
    \end{IEEEproof}
    
    The Lemma \ref{lemma2} reveals that the nuisance parameters and the time delay parameters are uncorrelated. 
    That is, the presence of nuisance parameters does not affect the CRLB of ToA-based target localization.

    According to the equations
    $\eqref{CRLB}, \eqref{Jphi}, \eqref{G}$, and $\eqref{Jpsi}$, 
    we can easily derive the sum of  CRLB for the ToA-based target localization, as shown in $\eqref{CRLBxy}$.
\end{IEEEproof}

\subsection{Problem Formulation}
Adopting the derived  communication and sensing performance metrics
in \eqref{eq-sinr} and \eqref{CRLBxy},
our primary task is to optimize the 
beamforming vectors $\left\{ \mathbf{f}_{m,k} \right\}$ to maximize the localization accuracy while
maintaining the communication performance. 
Mathematically, the optimization problem 
is formulated as
\begin{subequations}
    \begin{align}
        \mathop{\text{minimize }}\limits_{\left\{\mathbf{f}_{m, k}\right\}} & C_{x, y} \\
        \text {subject to }  & \sum_{k=1}^{K} \left\|\mathbf{f}_{m, k}\right\|^{2} \leq P, \ \forall m,  \label{p3} \\
        &\mathrm{SINR}_{m, k}  \geq \Gamma, \ \forall m, k \label{p2},
    \end{align} \label{eq:problem}%
\end{subequations}
where  $\Gamma$ is the SINR threshold.
It is imperative to note that the optimization problem in \eqref{eq:problem} 
exhibits highly 
nonconvex characteristics, especially due to the mixture of fractional structure and quadratic terms in the objective function,
which renders the problem challenging to solve by conventional methods.

%% file: subText/digital.tex
\section{Coordinated ISAC Beamforming}
In this section, we propose a generally global optimal solution to the optimization problem in \eqref{eq:problem}.
The non-convex nature of $C_{x,y}$ with respect to $\left\{\mathbf{f}_{m,k}\right\}$, 
stemming from its fractional structure and the presence of quadratic terms, 
complicates the resolution of problem \eqref{eq:problem}.
To tackle this challenge, we initially transform  
the fractional structure into two matrix inequalities, 
followed by employing the  semi-definite relaxation (SDR)
strategy.

Bearing this approach in mind, we initially introduce two auxiliary optimization 
variables, namely $\mu_1$ and $\mu_2$. 
Subsequently, by Schur's complement theorem, 
we   reformulate the problem in \eqref{eq:problem} into following equivalent form
\begin{subequations}
	\begin{align}
        \mathop{\text{minimize }}\limits_{{\left\{\mathbf{f}_{m, k}\right\}, \mu_1, \mu_2}} & \mu_1+\mu_2 \\
        \text {subject to } \ &  \eqref{p3}, \  \eqref{p2},  \\
		& \left[\begin{array}{cc}
			\mathbf{\Lambda Z} \mathbf{\Lambda}^T & \mathbf{e}_i \\
			\mathbf{e}_i^T & \mu_i
				\end{array}\right] \succeq \mathbf{0}, \ i=1,2, \label{nlmi}
    \end{align} \label{eq:problem_lmi}%
\end{subequations}
where $\mathbf{e}_i$ denotes the $i$-th column vector of the identity matrix of dimension 2.

Although  problem \eqref{eq:problem_lmi} remains nonconvex owing to the   constraints  in \eqref{p2}
and   \eqref{nlmi}, 
employing the SDR strategy allows for its relaxation into a convex formulation.
Specifically, 
by defining  $\mathbf{F}_{m,k} = \mathbf{f}_{m,k}\mathbf{f}_{m,k}^H, \forall m,k$, 
the equation \eqref{J_ele} can be reformulated  as 
\begin{equation}
    \begin{aligned}
        J(\tau_{m,n},\tau_{m,n}) & = \frac{8\pi^2 L T_{\mathrm{s}} \beta^2 \left| \varepsilon_{m,n} \right|^2}{\sigma_{\mathrm{s}}^2}
		\\ & \times  \mathbf{a}^T\left(\theta_m\right) 
        \left( 
        \sum_{k=1}^{K}\mathbf{F}_{m,k} \right)  \mathbf{a}^{*}\left(\theta_m\right).
    \end{aligned}  \label{J_ele4}
\end{equation}
Through this reformulation, 
constraint \eqref{nlmi} is transformed into a linear matrix inequality (LMI), 
establishing a convex constraint for $\left\{\mathbf{F}_{m,k}\right\}$. 
Nonetheless, the requirement for $\left\{\mathbf{F}_{m,k}\right\}$ to be of rank one introduces a new 
nonconvex constraint. To address this issue, we omit the rank-one constraint 
for $\left\{\mathbf{F}_{m,k}\right\}$, thereby relaxing the problem \eqref{eq:problem_lmi} to 
a more tractable form as
\begin{subequations}
    \begin{align}
        \underset{{\left\{\mathbf{F}_{m, k}\right\}, \mu_1, \mu_2}}{\text{minimize }} & \mu_1+\mu_2 \\
        \text {subject to } \ &\operatorname{tr}\left(\sum_{k=1}^K \mathbf{F}_{m, k} \right)
		\leq P, \ \forall m, \label{power2}  \\
        &              \begin{aligned} 
							& \Gamma \left( \sum_{(i,j) \neq (m,k)}   \mathbf{h}_{i, m, k}^H 
							\mathbf{F}_{i, j} \mathbf{h}_{i, m, k} +
							\sigma_{\mathrm{n}}^2 \right) \\
							& \leq \mathbf{h}_{m, m, k}^{H} \mathbf{F}_{m, k} \mathbf{h}_{m, m, k}, \ \forall m, k,
						\end{aligned} \label{sinr} \\
		& \left[\begin{array}{cc}
			\mathbf{\Lambda Z} \mathbf{\Lambda}^T & \mathbf{e}_i \\
			\mathbf{e}_i^T & \mu_i
		\end{array}\right] \succeq \mathbf{0}, \ i=1,2,  \label{lmis}\\
		&  \mathbf{F}_{m,k} \succeq \mathbf{0}, \ \forall m, k. \label{semid}
    \end{align}
    \label{eq:problem_eq}%
\end{subequations}
This is a convex SDP problem and can be solved in polynomial time.

The relaxation of  rank-one constraint for $\left\{\mathbf{F}_{m,k}\right\}$
may lead to solutions that are not rank-one,
which may result in sub-optimal solutions for the original problem \eqref{eq:problem}. 
Fortunately, 
the optimal solutions of 
problem  \eqref{eq:problem_eq}
are   always   rank-one under mild conditions.
This is stated in the following Theorem.
\begin{theorem}
	Supposed that problem \eqref{eq:problem_eq} is feasible.
	If $\mathbf{a}^*\left(\theta_m\right) \notin \mathrm{span}
	\left( \bigcup_{i,j} \mathbf{h}_{m,i,j} \right), \forall m$, 
	the optimal solutions $\left\{\mathbf{F}_{m,k}^{\star}\right\}$
	of the problem in \eqref{eq:problem_eq}
	are always rank-one.
\end{theorem}

\begin{IEEEproof}
	Owing to the page limit, the proof is omitted.
\end{IEEEproof}

The condition that $\mathbf{a}^*\left(\theta_m\right) \notin \mathrm{span}
\left( \bigcup_{i,j} \mathbf{h}_{m,i,j} \right), \forall m$ 
is almost always satisfied when $N_{\mathrm{t}}>MK$ due to the random nature of the channel vectors.
If encountering a
special case that
the  optimal solutions  $\left\{\mathbf{F}_{m,k}^{\star}\right\}$ are not  rank-one, then 
the Gaussian randomization method \cite{Luo2006}  can be employed
to extract the beamforming vectors $\left\{\mathbf{f}_{m,k}^{\star}\right\}$ 
from $\left\{\mathbf{F}_{m,k}^{\star}\right\}$.

%% file: subText/simulation.tex
\section{Numerical Simulations}
\label{sec:simulation}

\subsection{Simulation Setup and Benchmark Algorithms}
In the simulations, we consider a multi-cell ISAC system 
comprising $M=2$ BSs,  $N=4$ TMTs, and $K=4$ users per cell.  
The multi-cell ISAC system is deployed in a $200 \times 400$  m$^2$ area.
The BSs are positioned at coordinates $(80,80\sqrt{3})$m and $(80,-80\sqrt{3})$m, respectively. 
The target is located at the origin. 
The TMTs are positioned at $(30,30)$m, $(30,-30)$m, $(-30,30)$m, and $(-30,-30)$m, respectively. 
The coordinates of CUs within the area are generated randomly.

In the simulations, the channel coefficient $\varepsilon_{m, n}$ is 
defined as $\sqrt{F_{m,n}}\zeta_{m,n}$, where $F_{m,n} = \frac{\mathrm{c}^2}{f_{\mathrm{c}}^2(4\pi)^3}
\frac{1}{d_{m}^2 \left(d^{\prime}_{n}\right)^2}$ signifies 
the large-scale channel fading coefficient between the $m$-th BS and the $n$-th TMT. 
The parameters $d_m$ and $d^{\prime}_n$ denote the 
distances from the $m$-th BS and the $n$-th TMT to the target, respectively. 
Moreover, $f_{\mathrm{c}}$ represents the carrier frequency.
$\zeta_{m,n}$ indicates the RCS related to the $m$-th BS and the $n$-th TMT,
which is assumed to be a random variable following a Gaussian distribution with zero mean and variance of 1 in the simulations. 
The SINR threshold $\Gamma$ is set to 20 dB, and the transmit power $P$ is set to 30 dBm.
Additional simulation parameters are detailed in Table \ref{tab:simulation_parameters}.
\begin{table}[t]
	\centering
	\caption{Simulation Parameters}
	\begin{tabular}{|c|c|c|}
		\hline
		\textbf{Notation} & \textbf{Value} & \textbf{Description} \\ \hline
		$f_{\mathrm{c}}$ & 24GHz \cite{SturmWaveformDesignSignal2011} & Frequency of carrier \\ \hline
		$N_{\mathrm{t}}$ & 16 & Number of antennas at each BS \\ \hline
		$\beta$ & 100 MHz \cite{SturmWaveformDesignSignal2011} & Effective bandwidth \\ \hline
		$\sigma_{\mathrm{n}}^2$ & -94 dBm & Power of communication noise\\ \hline
		$\sigma_{\mathrm{s}}^2$ & -174 dBm/Hz \cite{huangCoordinatedPowerControl2022} & PSD of sensing noise\\ \hline
		$L$ & 256 \cite{SturmWaveformDesignSignal2011} & Number of sensing snapshot \\ \hline
	\end{tabular}
	\label{tab:simulation_parameters}
\end{table}

We compare the proposed beamforming algorithm with the radar only beamforming,
zero-forcing (ZF) beamforming \cite{chengOptimalCoordinatedTransmit2024},
and beampattern approximation beamforming \cite{liuJointTransmitBeamforming2020}.
	
\subsection{Performance Evaluation}
\begin{figure}
	\centering
	\includegraphics[width=0.45\textwidth]{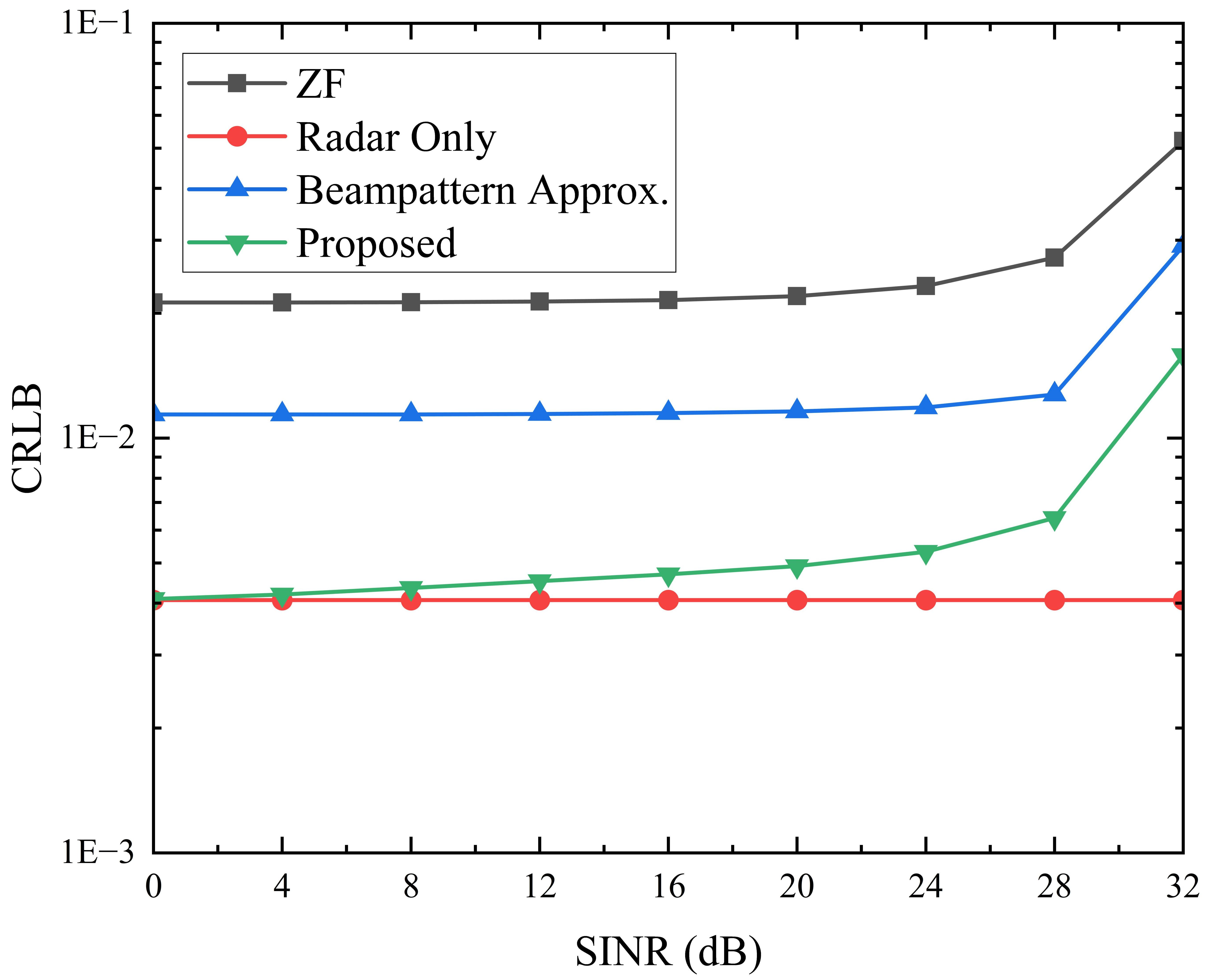}
	\caption{CRLB performance versus the SINR threshold $\Gamma$.}
	\label{figCRLBvsSINR}
\end{figure}

We firstly evaluate the CRLB performance versus the SINR threshold $\Gamma$, 
as illustrated in Fig. \ref{figCRLBvsSINR}. 
The results reveal that, irrespective of the SINR threshold, 
the proposed beamforming algorithm exhibit superior performance compared to both the ZF and 
beampattern approximation algorithms. 
Moreover, an increase in the SINR threshold   leads 
to a corresponding rise in the CRLB for the proposed algorithm. 
This trend indicates the inherent performance 
trade-off between 
target localization accuracy and communication rate.

\begin{figure}
	\centering
	\includegraphics[width=0.45\textwidth]{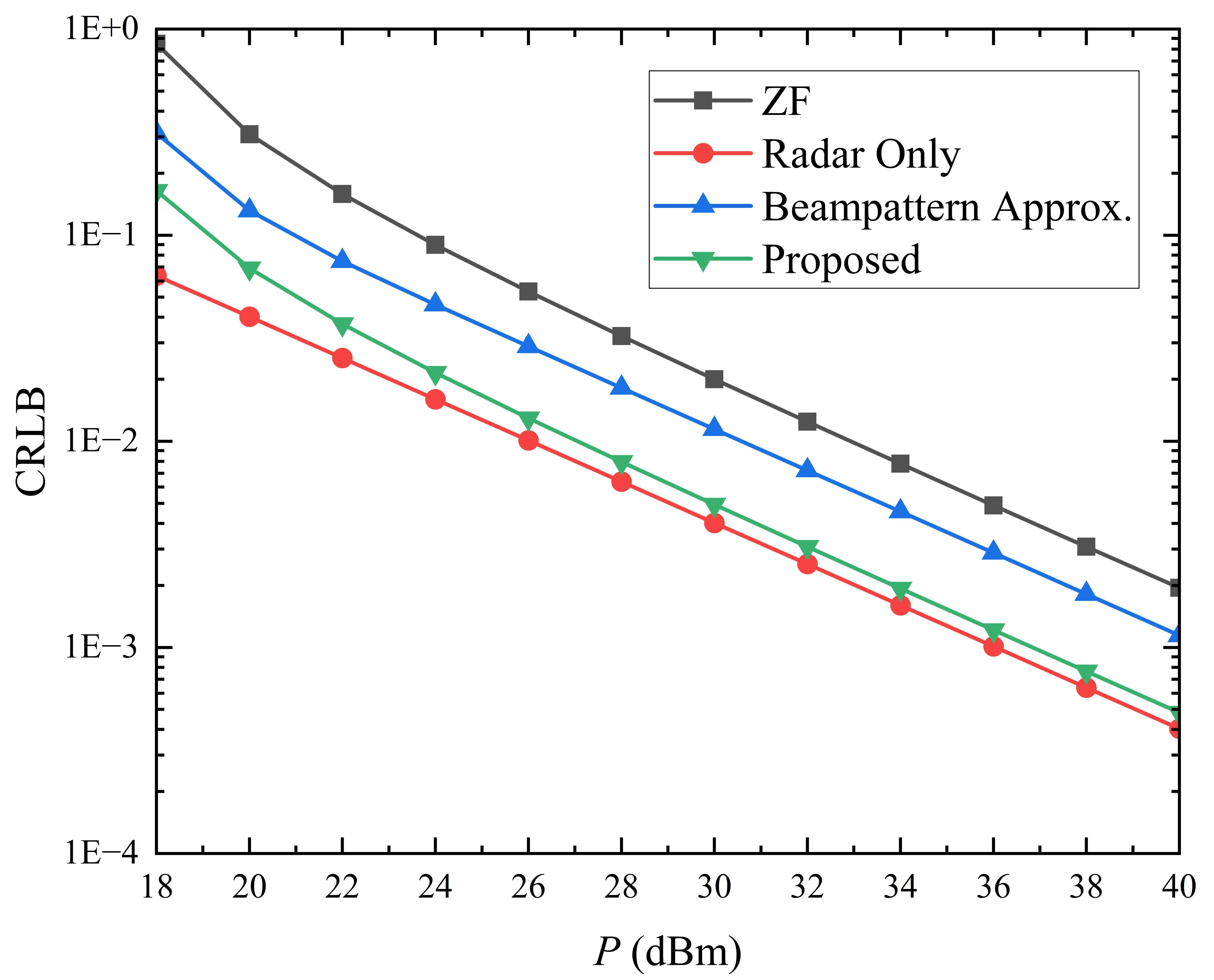}
	\caption{CRLB performance versus the   transmit power $P$.}
	\label{figCRLBvsP}
\end{figure}

Fig. \ref{figCRLBvsP} compares the CRLB performance versus the  transmit power $P$. 
The proposed algorithm
consistently surpasses the performance of both the ZF and beampattern approximation algorithms, 
irrespective of the magnitude of $P$.
Furthermore, an intriguing trend is observed regarding the performance disparity 
between the proposed algorithm and the radar only beamforming algorithm. 
Specifically, this gap widens as $P$ diminishes. 
The underlying reason is that the ISAC system must 
allocate a sufficient portion of the   transmit power to   communication  
to ensure the maintenance of the communication rate. Consequently, this necessitates 
a reduction in the power allocated to the sensing   when $P$ is constrained, 
highlighting the critical trade-off between target localization accuracy and communication SINR.

\begin{figure}
	\centering
	\includegraphics[width=0.45\textwidth]{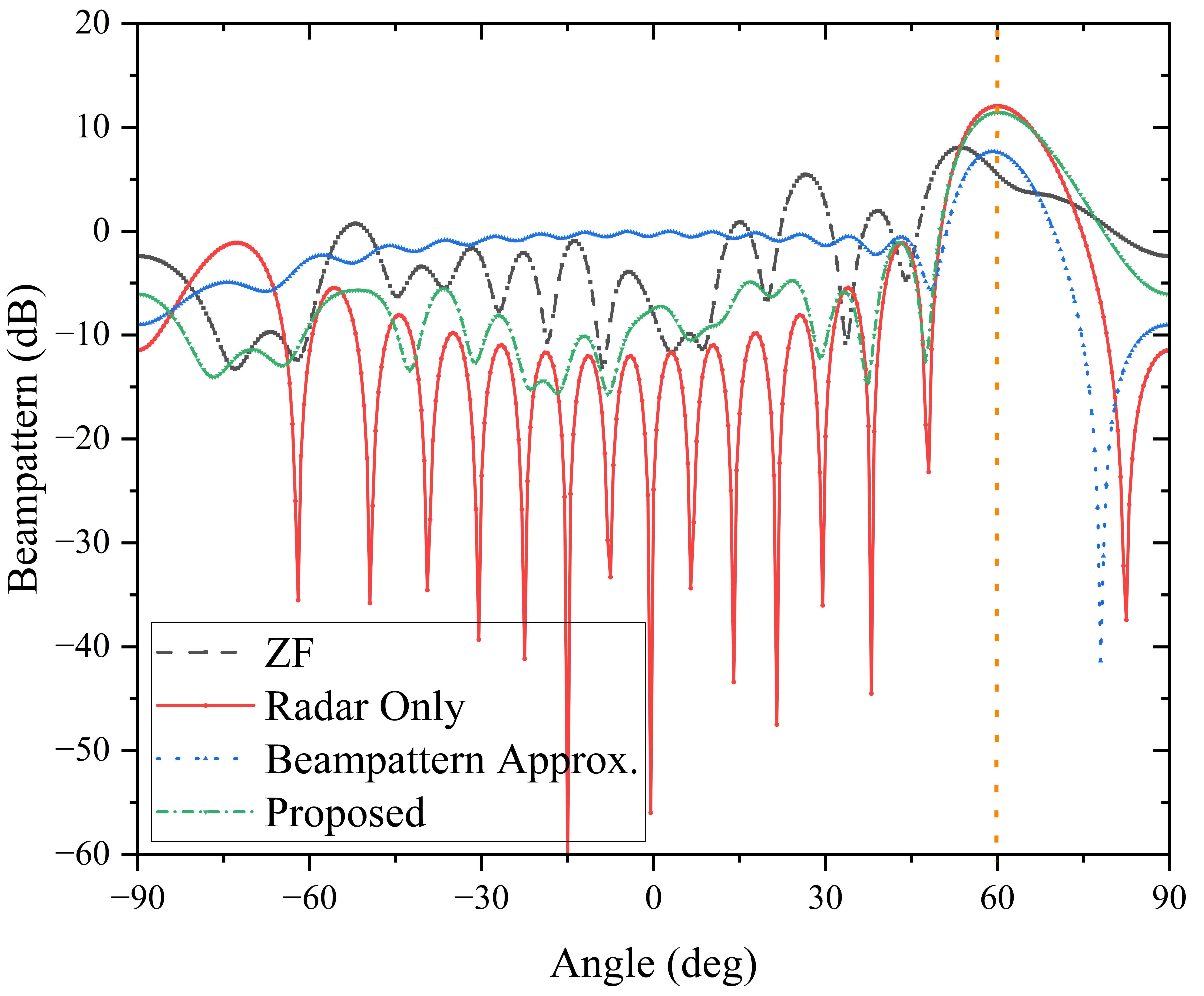}
	\caption{Beampatterns for all considered algorithms in BS 1.}
	\label{figbeampattern}
\end{figure}

The resultant beampatterns for all  considered algorithms in BS 1 are depicted in Fig. \ref{figbeampattern}.
Within this simulation, the target is positioned at angle of $60^\circ$  relative 
to the  BS.  
Observations reveal that the proposed beamforming algorithm successfully generate 
the desired beampattern, with its main lobe   aligned at $60^\circ$, 
mirroring the performance of the radar only beamforming algorithm. 
However, the beampattern approximation algorithm does not achieve the 
similar  
level of sharpness in the main lobe as the proposed algorithm.
Moreover, the ZF algorithm fails to form directional beams 
towards the target direction, 
underscoring its inadequacy for radar sensing applications.
These simulation outcomes align with the CRLB performance results presented in 
Fig. \ref{figCRLBvsSINR} and Fig. \ref{figCRLBvsP}, 
further validating the efficiency and suitability of the proposed beamforming 
algorithm for multi-cell integrated localization  and communication systems.

%% file: subText/Conclusion.tex
\section{Conclusion}
In this paper, we studied the integrated localization and communication problem in a
multi-cell system.
We derived two key performance indicators: the SINR for communication and the CRLB
for target localization.
Exploiting the established 
performance metrics, we formulated a nonconvex optimization problem to minimize the CRLB
while maintaining the communication SINR.
We proposed a coordinated beamforming algorithm to solve this problem based on the SDR strategy,
and   show that the SDR is always tight under mild conditions.
Through   numerical simulations, we verified the effectiveness of the proposed algorithm, 
and show the trade-off between communication SINR and  the
CRLB of target localization, providing 
valuable insights for the design of future multi-cell ISAC systems.
\addtolength{\topmargin}{0.05in}

%% file: subText/Appendix.tex
\appendices
\section{Proof of Lemma 2}
The $(l,p)$-th element of the FIM $\mathbf{J}(\boldsymbol{\uppsi})$ is determined through
the Slepian-Bang formula \cite{stoicaSpectralAnalysisSignals2005} 
\begin{equation}
    J(\psi_p,\psi_l)=\frac{2}{\sigma_{\mathrm{s}}^2} 
    \int_{0}^{LT_{\mathrm{s}}}\operatorname{Re}\left( \frac{\partial \boldsymbol{\upmu}^H(t)}{\partial \psi_l} 
    \frac{\partial \boldsymbol{\upmu}(t)}{\partial \psi_p}\right) \mathrm{d}t, \ \forall l, p \label{26},
\end{equation}
where $\boldsymbol{\upmu}(t)=\left[\mu_1(t),\mu_2(t),\ldots,
\mu_N(t)\right]^T$. 

Following equation \eqref{26}, we derive \cite{lehmannHighResolutionCapabilities2006}
\begin{equation}
    \begin{aligned}
        & J(\tau_{m,n},\tau_{m^{\prime},n^{\prime}}) = 
        \left\{ \begin{array}[2]{ll}
            \frac{2}{\sigma_{\mathrm{s}}^2}    u_{m,n}     , & m=m^{\prime},n=n^{\prime} \\
            0, & \text{otherwise}
        \end{array} \right. ,
    \end{aligned} \label{1}
\end{equation}
where 
\begin{equation}
    \begin{aligned}
        u_{m,n} & = 4\pi^2 L T_{\mathrm{s}} \beta^2 \left| \varepsilon_{m,n} \right|^2      \\
        & \times \mathbf{a}^T\left(\theta_m\right) 
        \left( 
        \sum_{k=1}^{K}\mathbf{f}_{m,k} \mathbf{f}_{m,k}^{H} \right)  \mathbf{a}^{*}\left(\theta_m\right),  
    \end{aligned}
\end{equation}
and $\beta   \triangleq \sqrt{\frac{\int_{-\infty}^{\infty} f^2 \left|G(f)\right|^2 \mathrm{d}f}
{\int_{-\infty}^{\infty} \left|G(f)\right|^2 \mathrm{d}f} }$.
Similarly, we establish \cite{lehmannHighResolutionCapabilities2006} 
\begin{subequations}
    \begin{align}
        J(\tau_{m,n},\theta_{m^{\prime}}) &= 0, \  \forall m, n,  m^{\prime},  \label{22} \\
        J(\tau_{m,n},\varepsilon_{m^{\prime},n^{\prime},\mathrm{R}}) &= 0, \ \forall m, n, m^{\prime}, n^{\prime},  \label{3} \\
        J(\tau_{m,n},\varepsilon_{m^{\prime},n^{\prime},\mathrm{I}}) &= 0, \ \forall m, n, m^{\prime}, n^{\prime}, \label{44}
    \end{align}
\end{subequations}
where $\varepsilon_{m^{\prime},n^{\prime},\mathrm{R}}$ and
$\varepsilon_{m^{\prime},n^{\prime},\mathrm{I}}$ represent the real 
and imaginary part of $\varepsilon_{m^{\prime},n^{\prime}}$, respectively.
From equation \eqref{22} to equation \eqref{44}, we have used the equation \eqref{eq:real} \cite{lehmannHighResolutionCapabilities2006}.

Based on    equations \eqref{22}, \eqref{3} and \eqref{44}, 
we can deduce 
that the nuisance parameters and the time delay parameters are uncorrelated.
Furthermore, it is clear that the FIM $\mathbf{J}(\boldsymbol{\uppsi})$
is a block-diagonal matrix  as denoted by
\eqref{Z},  \eqref{J_ele}, and \eqref{Jpsi}.
